# Finding beam loss locations in a linac with oscillating dipole correctors

A. Shemyakin, Fermilab, Batavia, IL 60510, USA
*shemyakin@fnal.gov*

*Abstract*
The paper proposes a method of finding the beam loss locations in a linac. If the beam is scraped at an aperture limitation, moving its centroid with two dipole correctors located upstream and oscillating in sync produces a line at the corresponding frequency in spectra of current-sensitive devices downstream of the loss point. The phase of this signal contains information about the location of the beam loss. Similar lines appear also in the position signals of Beam Position Monitors (BPMs). The phases of the BPM position lines change monotonically (within each $2\pi$) along the linac and can be used a reference system. The phase of the loss signal compared with this reference system pinpoints the beam loss location, assuming that longitudinal coordinates of the BPMs are known.
If the correctors deflection amplitudes and the phase offset between their waveforms are chosen optimally and well calibrated, the same measurement provides values of the β-function and the betatron phase advance at the BPM locations. Optics measurements of this type can be made parasitically, with negligible effect on the emittance, if a long measurement time is acceptable.

# Table of Contents

# 1. Introduction

A usual requirement for a linear accelerator (linac) or a beam line[1] is to transport the beam with low losses to avoid excessive residual radiation and degradation of equipment. A standard way to identify the beam loss is to measure the radiation produced by the lost particles or to compare reading of beam current measuring devices in various locations. Both methods have certain limitations.

Radiation monitoring do not work effectively at low particle energies in the case of ion beams. Also, the monitors need to be placed not too far the loss location, leaving a chance that some loss places are missed. On the other hand, in the case of high radiation, a monitor signal might be sensitive to losses over a comparatively large length of the linac, not allowing to accurately locate the loss point. Finally, sometimes the beam-induced radiation needs to be distinguished from other sources of radiation, e.g. from RF resonators.

So-called Differential Beam Current Monitor (DBCM) system that compares currents from two or more current-reading devices is successfully used for detection of a large accidental loss [1], but sensitivity to low, long-lasting losses is limited to percent level by resolution of the current measuring devices and their noise. Also, the DBCM does not reveal the actual location of the loss.

When the approximate location of the beam loss is known, the typical way to find and address an "aperture limitation"[2] is to create a localized perturbation and check whether it can decisively affect the loss. For example, three dipole correctors can create a "three-bump" trajectory perturbation shifting the trajectory with maximum deviation at the location of the central corrector while leaving the beam unchanged outside of the up- and downstream correctors.

This paper proposes a procedure to both identify existence of a small beam loss and find its location. For this purpose, currents in a couple of dipole correctors positioned upstream of the expected loss location (e.g. at the beginning of the linac) are oscillated in sync at a fixed frequency. If the beam is being scraped, the oscitation results in a variation of the beam current at this frequency that can be observed in the signals of the linac diagnostics. If, in addition, oscillations in the transverse beam positions are measured, one can reconstruct the transverse optics of the accelerator.

The proposal is originated from a discussion with Valeri Lebedev [2] on detection of small losses in a CW ion accelerator. Similar measurements were introduced by him in CEBAF at the end of nineties. Valeri pointed out that, in practice, a beam loss caused by beam scraping on a physical aperture is always associated with a sharp dependence of the loss signal on the beam position at the aperture limitation. Therefore, if an aperture limitation exists, oscillating a dipole corrector current at the beginning of the linac is likely to produce the corresponding line in the beam current signal measured at the end of the linac. Such measurement does not provide an absolute value of the loss but rather the difference in the loss over the range of the beam oscillation. While typically the latter is lower than the total loss, the synchronous detection allows to greatly improve the overall sensitivity. For sufficiently long measurement time, the oscillations resulting in a detectable effect can be made small enough so that the effective emittance growth is negligible (i.e. the beam phase space footprint does not change any significantly). What follows below is development of this idea.

---

[1] Since this paper considers only transverse motion, the results are equally applicable to a beam line with constant energy and a linac, so that in this paper the terms are used interchangeably.

[2] When the beam loss is sensitive to many settings upstream, for a machine operator it feels like the beam is passing through a narrow opening. While in practice a much more common case is non-optimum focusing or steering settings, an "aperture limitation" is an easily-to-visualize way to describe a localized, steady beam loss.

## 2. Simple model with a single oscillating dipole corrector

Let's consider the simplest model of a one-dimensional transverse motion of the beam excited by a single dipole corrector in a channel with constant focusing over one betatron period of length $\lambda$ (Fig.1). The beam has a one- dimensional current density distribution $j(x-x_0)$ independent on the longitudinal coordinate $z$, where $x_0$ denotes the transverse position of the central trajectory. A dipole corrector at the beginning of the channel deflects the beam by the angle $\theta$, and the central particle trajectory is

$$x_0(z) = \frac{\theta}{k}\sin kz, \tag{1}$$

where $k = 2\pi/\lambda$.

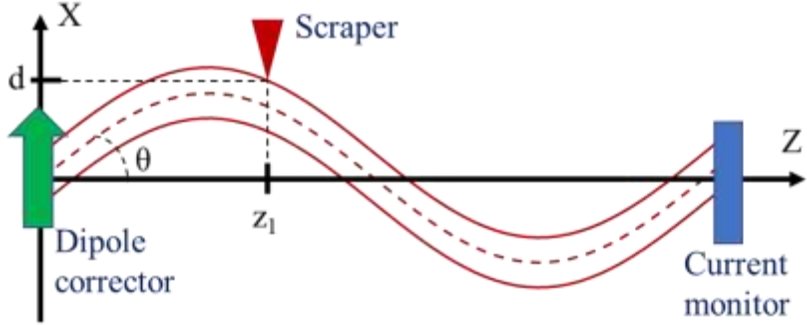

Figure 1. Simple model with one oscillating dipole corrector.

An aperture limitation is approximated by a flat-edge "scraper" that is located at $z_1$ and separated from the axis by $d$. The scraper intercepts a part of the beam current

$$I_s = \int_d^\infty j(x - x_0)dx. \tag{2}$$

The interception results in a deviation of the beam current measured by a monitor at the end of the line, $I_m$, from its initial value $I_0$, $I_m = I_0 - I_s$

Now consider the case when the dipole corrector current oscillates in time,

$$\theta(t) = \theta_0 \sin \omega t. \tag{3}$$

The Fourier spectrum of the current monitor will exhibit harmonics of $\omega$. In the case of a smooth function $j(x)$, Eq.(2) can be expanded around $x_0 = 0$:

$$I_s = \int_d^\infty j(x)dx + j(d)x_0 - \frac{dj}{dx}(d)\frac{x_0^2}{2} + \cdots \tag{4}$$

For small oscillation amplitudes, only the first terms are significant,

$$I_s \approx \int_d^\infty j(x)dx + j(d)\frac{\theta_0}{k}\sin \omega t \sin kz_1 - \frac{dj}{dx}(d)\frac{1}{2}\left(\frac{\theta_0}{k}\sin \omega t \sin kz_1\right)^2 \equiv$$

$$\equiv I_{s0} + I_{s1}\sin \omega t + I_{s2}(1 - \cos 2\omega t). \tag{5}$$

The ratio of amplitudes of the second and first harmonics is

$$\frac{I_{s2}}{I_{s1}} = -\frac{\frac{dj}{dx}(d)}{4\cdot j(d)}\left(\frac{\theta_0}{k}\sin kz_1\right). \tag{6}$$

For example, if the beam distribution is Gaussian,

$$j(x) = j_G(x) \equiv \frac{1}{\sqrt{2\pi}\sigma} e^{-\frac{x^2}{2\sigma^2}}, \tag{7}$$

this ratio is

$$\frac{I_{s2}}{I_{s1}} = \frac{d}{4\cdot\sigma^2}\frac{\theta_0}{k}\sin kz_1 . \tag{8}$$

In the interesting range of parameters, the first harmonic dominates. For example, if oscillation amplitude at the scraper location is $0.1\sigma$ and the scraper is at $d/\sigma = 2$ (intercepting ~2% of the beam), the amplitude of the second harmonic is only 5% of the first one. The amplitude of changes in the beam loss is $I_{s1}/I_0 = 0.5\%$, is significantly lower than the total loss.

In some cases, the second harmonic may be important. For example, if the beam is scraped on both sides simultaneously and is well aligned, the first harmonic is suppressed, and the loss signal will be defined by the second harmonic. However, for simplicity, this paper considers only the first harmonic, i.e. the signals at the frequency of the oscillating dipole correctors, which will be referred as an oscillation line (in the Fourier spectrum). Note that by limiting the signal to its linear component in Eq.(5), the procedure, in a sense, measures the derivative of the beam loss over the transverse beam position.

The scheme has a clear drawback: the aperture limitation is undetectable if the scraper location is close to $z_1 = \lambda/2 \, n$. Also, it can determine that a loss occurs between the dipole corrector and the current monitor but cannot pinpoint the location more accurately. These issues can be addressed by using two dipole correctors.

## 3. Simple model with two oscillating dipole correctors

Let's append the model of Fig.1 with an additional dipole corrector placed by $\lambda/4$ upstream of the first one (Fig.2).

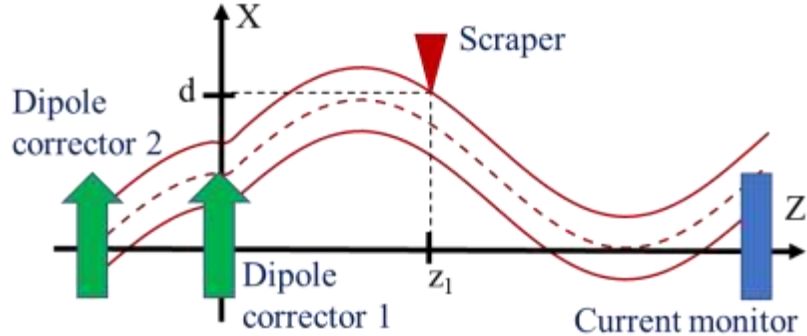

Figure 2. Simple model with two oscillating dipole correctors.

This second corrector can probe all locations unreachable for the first one. If both correctors oscillate with the same amplitude and at the same frequency but shifted in time by a quarter of the period, the trajectory transforms from a standing wave to a travelling wave:

$$x_0(z,t) = \frac{\theta_0}{k}\sin kz \, \sin \omega t + \frac{\theta_0}{k}\sin\left(kz + \frac{\pi}{2}\right)\sin\left(\omega t + \frac{\pi}{2}\right) = \frac{\theta_0}{k}\cos(\omega t - kz) . \tag{9}$$

Therefore, during a full oscillation period the beam is shifted by the same amount in all locations in the beam line. The specific moments of time when the shift is maximum are defined by the longitudinal coordinate. Correspondingly, the loss signal contains information about the location of the loss in its phase:

$$I_s(t) \approx I_{s0} + j(d)\frac{\theta_0}{k}\cos(\omega t - kz_1). \tag{10}$$

The synchronous detection (or Fourier analysis of the recorded loss signal) allows to determine the phase of the loss with respect to the dipole corrector oscillation and from Eq.(10) the location of the loss with uncertainty of $\pi n$ (taking into account that the beam can be scraped on either side).

## 4. A realistic model

In a real accelerator, the linear, one-plane uncoupled beam motion is described by Twiss functions $\beta_x, \alpha_x$. Two dipole correctors separated by the betatron phase advance $\varphi_x$ and oscillating at the same frequency with the phase offset $\varphi_t$ shift the central particle by

$$x_0(z,t) = \theta_1\sqrt{\beta_x(z)\beta_{x1}}\sin\varphi(z)\,\sin\omega t + \theta_2\sqrt{\beta_x(z)\beta_{x2}}\sin(\varphi(z) + \varphi_x)\sin(\omega t + \varphi_t). \tag{11}$$

where indexes 1 and 2 mark the values for corresponding correctors and $\varphi(z)$ is the betatron phase advance in the beam line counted from the corrector 1.

In all locations, the beam position oscillates at the same frequency but with amplitude and phase dependent on $z$:

$$x_0(z,t) = A(z)\sin(\omega t + \varphi_1(z)) \tag{12}$$

Transformation of Eq. (11) gives

$$\left[\frac{A(z)}{\theta_1\sqrt{\beta_x(z)\beta_{x1}}}\right]^2 = (1 + a\cos\varphi_x\cos\varphi_t) + \sin 2\varphi(z)\cdot a\sin\varphi_x(\cos\varphi_t + a\cos\varphi_x) -$$
$$- \cos 2\varphi(z)\cdot\left[\frac{1-a^2}{2} + a\cos\varphi_x(\cos\varphi_t + a\cos\varphi_x)\right], \tag{13}$$

where

$$a \equiv \frac{\theta_2\sqrt{\beta_{x2}}}{\theta_1\sqrt{\beta_{x1}}}\,. \tag{14}$$

If the particle distribution in action – phase coordinates is not phase-dependent and stays constant during propagation, the spatial beam distribution is scaled along the beam line as the beam rms size $\sigma_b = \sqrt{\beta_x(z)\varepsilon_0}$, where $\varepsilon_0$ is the rms beam emittance. In this approximation, the current density distribution can be written as

$$j(x) = \frac{I_0}{\sigma_b}\widetilde{J}\left(\frac{x}{\sigma_b}\right), \tag{15}$$

where the dimensionless function $\widetilde{J}\left(\frac{x}{\sigma_b}\right)$ doesn't change along the beam line. Correspondingly, the amplitude of the loss linear component is determined by the scraper position and oscillation amplitude, both normalized by the rms beam size:

$$\frac{I_{s1}}{I_0} = \frac{A(z)}{\sigma_b}\widetilde{J}\left(\frac{d}{\sigma_b}\right). \tag{16}$$

In the case when the right side of Eq.(13) does not depend on $z$, the oscillation amplitude changes along the beam line as $\sqrt{\beta_x(z)}$, and the procedure probes evenly all location. However, generally the relative amplitude does change with the phase advance at the double betatron frequency. Rewriting Eq.(13), the expression for the relative amplitude is

$$\frac{A(z)}{\sigma_b} = \theta_1 \sqrt{\frac{\beta_{x1}}{\varepsilon_0}} [(1 + a \cos\varphi_x \cos\varphi_t) + B \sin(2\varphi(z) + \varphi_2)]^{1/2},$$
$$B^2 = \left(\frac{1-a^2}{2}\right)^2 + a(1 + a^2) \cos\varphi_x \cos\varphi_t + a(\cos\varphi_t^2 + \cos\varphi_x^2). \quad (17)$$

The relative amplitude becomes position-independent when $B = 0$. For that, one needs, first, to adjust the deflection amplitudes by the dipole correctors so that $a = 1$ or

$$\theta_2 \sqrt{\beta_{x2}} = \theta_1 \sqrt{\beta_{x1}} \quad (18)$$

and, second, the phase shift in time between them so that $\cos\varphi_t = -\cos\varphi_x$, or

$$\varphi_t = \pi - \varphi_x. \quad (19)$$

When conditions Eq.(18) and Eq.(19) are fulfilled, Eq.(12) becomes

$$x_0(z,t) = \theta_1 \sqrt{\beta_x(z)\beta_{x1}} \sin\varphi_x \sin(\omega t + \varphi_1(z)) \quad (20)$$

with the phase shift, as can be shown from analyzing Eq.(11) and Eq.(12),

$$\varphi_1(z) = -\varphi(z) - \varphi_x. \quad (21)$$

The case with the dipole correctors separated by exactly 90° of the betatron phase advance,

$$\varphi_x = \pi/2, \quad (22)$$

is identical to the one considered in Section 3 and described by Eq.(10): at the initial plane ($\varphi = 0$), corrector 1 defines the angle, and corrector 2 defines the initial position.

Equations (20) and (21) show that if
- there are two dipole correctors with the betatron phase being not far from $\pi(n + \frac{1}{2})$;
- the optics is well known (specifically, the phase advance along the beam line and $\beta$-functions in the corrector locations are well determined);
- correctors deflections are calibrated;
- there is a current monitor downstream;

it allows to determine the loss location (within $\pi$ of the betatron phase advance) by
- setting up waveforms at the correctors to fulfill Eq.(18) and (19);
- recording the signal from the current monitor for long enough time to observe a clear spectrum line at the oscillation frequency above the noise;
- determining the loss coordinate from the known betatron phase along the beam line.

## 5. Case of a not-well-known optics

The procedure outlined above is not directly applicable when the actual optics is not known accurately. However, in a real beamline or linac, oscillating dipole correctors still can help to pinpoint the beam loss location if more diagnostics are used.

In modern machines, the beam position is measured by Beam Position Monitors (BPMs). Typically, they are capacitive pickups with electronics that delivers a differential signal between opposite plates converted to the transverse position and a sum signal proportional to the bunch intensity (referred here as a BPM intensity). If a couple of dipole correctors is set to oscillate in sync, the BPM positions exhibit the corresponding line in their Fourier spectrum. Obviously, the amplitude $A_i$ and phase $\varphi_i$ of this oscillatory term contain information about the beam line optics. This information can be interpreted and used as follows.

Reasoning in the Section 4 assumes implicitly that the initial beam phase portrait is known, the initial Twiss parameters of the beam line are chosen to match the beam, and preparation of the dipole correctors waveforms to fulfill Eq. (18)-(19) is based on this choice. At arbitrary values of the ratio between the correctors deflections and the time shift between their waveforms, without any reliance on the beam envelope information, the logic of Section 4 can be reversed. The beam centroid position can still be described by Eq. (12), but with the initial Twiss parameters *defined* by the conditions of Eq. (18)-(19).

To make the statement clearer, note that the beam position and angle at the location of the corrector 1 depends on only two elements of the transition matrix between two dipole correctors since the trajectory starts at the corrector 2 with zero offset. These elements can be expressed through Twiss functions as follows (e.g. [3]):

$$m_{12} = \sqrt{\beta_{x2}\beta_{x1}} \sin \varphi_x;$$
$$m_{22} = \sqrt{\frac{\beta_{x2}}{\beta_{x1}}} (\cos \varphi_x - \alpha_{x1}\sin \varphi_x). \tag{23}$$

Considering now Eq. (18)-(19) as conditions on the β-functions and the phase advance for the given time phase and deflection amplitudes,

$$\sqrt{\frac{\beta_{x2}}{\beta_{x1}}} = \frac{\theta_2}{\theta_1}; \quad \varphi_x = \pi - \varphi_t, \tag{24}$$

one can find corresponding initial Twiss parameters

$$\beta_{x1} = \frac{m_{12}}{\sin \varphi_t}\frac{\theta_1}{\theta_2}; \quad \alpha_{x1} = -\frac{m_{22}\frac{\theta_2}{\theta_1}+\cos \varphi_t}{\sin \varphi_t}. \tag{25}$$

Initial conditions of Eq.(25) are well-defined most of the time with exception of obvious cases $\theta_1 = 0$, $\theta_2 = 0$, $\sin \varphi_t = 0$ as well as $m_{12} = 0$ when the upstream corrector (#2) does not create an offset in the location of the corrector #1, so that both correctors affect the trajectory identically.

Therefore, the oscillation phases and amplitudes measured by BPMs reflect, correspondingly, the betatron phases $\{\varphi_i\}$ and β-functions at BPM locations through Eq.(20)-(21):

$$A_i = \theta_1\sqrt{\beta_x(z_i)\beta_{x1}} \sin \varphi_t; \quad \varphi_i = -\varphi(z_i) + \varphi_t - \pi, \tag{26}$$

where $\{z_i\}$ are the BPM longitudinal coordinates. These phases and Twiss functions are defined with specific initial conditions of Eq.(25), which can differ from those for a matched beam.

For a large enough number of BPMs (several per one betatron period) with known locations, interpolation of the BPM betatron phases between $\{z_i\}$ creates a map of phases along the beam line. If the spectrum of a current monitor shows a line at the correctors' oscillation frequency, the phase of this line determines the location of the beam loss (within πn).

Additional type of diagnostics is radiation-sensitive Beam Loss Monitors (BLMs). BLMs report signals that are linear with the average beam loss, though amplitudes of their signals are affected by many parameters as the beam energy, distance between the BLM and loss location, etc. However, the phase of the oscillation line in the BLM signal delivers the information of the loss location in the same manner as a current monitor, independently on the mutual location of the BLM and the loss point.

Finally, the BPM intensities are also sensitive to the beam loss. Using them for loss identification directly, by comparison of absolute values of BPM intensities in neighboring BPMs, is difficult since they are sensitive to many factors as the bunch size, electronics calibration etc. However, all these factors are not likely to change during a single measurement with oscillating dipole correctors, and the amplitude of the oscillation line normalized by the average BPM intensity is a good measure of the beam loss derivative. Also, the phase of this line provides information about the loss location as in the case of a current monitor or a BLM. A significant benefit of using BPM intensities is a typically larger number of BPMs in comparison with current monitors, which eliminates the πn uncertainty. Resolution of these measurements is likely to be determined by quality of separation between differential (position) and sum (intensity) BPM signals, or, in other words, by how constant the intensity signal stays when the beam position changes.

## 6. Case of multiple loss points

When comparable beam losses occur in several locations at once, the oscillation line in a downstream current-sensing diagnostic (a current monitor or a BPM intensity) is a vector sum of losses upstream, so the line amplitude can go up or down in comparison with the previous element. If, for example, two identical aperture limitations are on the same side and separated by π of the betatron phase advance, the first harmonic is absent in the resulting current loss signal (though the second harmonic is present). However, in the case of multiple BPMs per betatron period, intensity signals in some of them would be different, indicating a loss point in between. Note that a BLM signal will be still sensitive to the closest loss location.

The procedure of locating multiple loss points can be as follows:
- Oscillate two dipole correctors that are upstream of the interesting area and downstream of known large unavoidable losses (e.g. intentional scraping). Record signals of all BPMs, BLMs, and current monitors;
- Perform the Fourier transform of BPM positions. Map phases of the line at the oscillation frequency to the longitudinal coordinate of the beam line using the known BPM locations;
- Normalize the oscillation signals of BPM intensities and current monitors by their corresponding average values to remove calibration effects;
- Calculate differences between these relative intensities of each BPM and the preceding one. Do the same for current monitors. Perform the Fourier transform of the differential signals and determine the amplitude and phase of the oscillation line.
- For the differential channels with large amplitudes of the oscillation line, compare the line phases with the phase map created by analysis of the BPM position signals and determine the loss locations. Do the same for current monitors;
- Perform the Fourier transform of BLM signals; compare finding with other results;
- A large beam loss found at the beginning of the beam line would dominate the current signals downstream, which can increase the noise in differential signals. If this loss is not easily correctable, the smaller losses further in the beam line can be found by repeating the procedure with another pair of dipole correctors downstream of the large loss.

Note that while this procedure is applicable to the case of arbitrary chosen oscillation parameters, knowing the beam envelope parameters in the location of the correctors is highly beneficial. If the waveforms are generated according to the recipe in Section 4 (Eq.(18)-(19)), the beam oscillates by the same portion of its rms size in all locations. As a result, the strength of the oscillation line in the differential intensity signals reflects the absolute beam loss (under assumption that Eq.(15) is valid). The further the oscillation amplitudes in the correctors and the time shift between them deviate from the optimum, the less certain is the relationship between the strength of the oscillation lines and absolute beam loss.

## 7. Notes on practical implementation

For practical realization of the proposed scheme, one can envision dedicated high-frequency deflectors to create the oscillating trajectory and a feature in the BPM and current monitor electronics performing synchronous detection at this frequency, as it was implemented by V. Lebedev at CEBAF. In this scenario, the scheme can be potentially used in the Machine Protection System to interrupt the beam comparatively quickly if a large enough amplitude signal is found in the current-sensitive devices.

The goal can be more modest, just to identify locations of losses that are small enough to do not create an immediate problem (like a hole in the vacuum pipe) but still significant to be a concern in a long run, for example, because of residual activation. In that case, the procedure can be implemented using existing equipment and a long acquisition time. In the first proof-of-principle measurements, detail description of which is outside of the scope of this paper, the acquisition time was ~10 min, the existing dipole correctors oscillated with 0.5-1 Hz, and the data were recorded every beam pulse at ~15 Hz with standard diagnostics. At this low oscillation frequency, the skin effect is negligible and corrector currents follow reasonably well to the settings even in the typical case of the correctors designed for a DC use only.

At higher oscillations frequencies, the beam is deflected at the same frequency, but the amplitude can decrease, and the phase can shift because of corrector power supplies' properties or eventually the skin effect. If the portion of the beam line immediately downstream of the correctors is optically known and contains BPMs, the dipole fields produced by the oscillating correctors can be calibrated. Since the required trajectory oscillations are small (much less than the beam rms size), increasing of the oscillation amplitudes is not a problem, and the setting waveforms can be properly modified in accordance with calibration measurements to generate the optimal deflection fields. Increasing the frequency might be beneficial when the noise spectral density decreases with frequency.

To apply such procedure parasitically, the associated emittance growth should be low. Corresponding estimation is given in Appendix and yields 2% emittance growth when the beam is shifted by 20% of its rms radius.

The procedure under discussion in this paper describes motion in one plane. Obviously, it can be applied sequentially to both planes. Alternatively, to sense also out-of-plane aperture restrictions, horizontal and vertical pairs of correctors can be oscillated simultaneously at well-separated frequencies. The out-of-plane restrictions should manifest themselves as signals at frequencies equal to sum and difference of horizontal and vertical oscillation frequencies.

## 8. Optics measurements

As it was mentioned in Section 5, the procedure delivers information about the optical properties of the line. In general case, it is a description with a particular choice of the initial Twiss parameters according to Eq. (25).

If the initial beam characteristics and correctors calibrations are known, and the correctors waveforms follow Eq. (18)-(19), the amplitudes and phases of BPMs readily deliver values of the $\beta$-functions and betatron phases according to Eq. (20)- (21), which can be used to refining the optics models. In optical sense, the results are equivalent to the response matrix in the method of differential trajectories (e.g. [4]). The benefits of obtaining them with oscillating dipole correctors come from the lower values of required variations in the beam positions: a possibility to take data parasitically as well as a lower sensitivity of the results to beam losses and magnets non-linearity.

## 9. Summary

To find a one-plane aperture limitation in a beam line, two dipole correctors are oscillated in sync and responses of downstream diagnostics (BPMs, current monitors, and BLMs) at the oscillation frequency are recorded. Phases of the BPM position signals interpolated over the known BPM longitudinal positions provide a reference system along the beam line. Comparing the phases of the oscillation line observed in the BPM intensity or in a current monitor (or in a differential signal between neighboring current-sensitive diagnostics in the case of multiple loss points) with the phase-based reference system allows to pinpoint the loss locations.

## 10. Acknowledgement

I am thankful to V. Lebedev for the original idea and discussions as well as for reading and suggesting corrections to this paper. The proposal to use two correctors was put in writing by Mantas Pastolis during his summer 2018 internship [5]. Useful discussions with Ram Prakash and Kiyomi Seiya during proof-of-principle tests of the method as well as K. Seiya's proposal to look at the oscillation line in the BLM signals are acknowledged and appreciated.

This manuscript has been authored by Fermi Research Alliance, LLC under Contract No. DE-AC02-07CH11359 with the U.S. Department of Energy, Office of Science, Office of High Energy Physics.

## 11. Appendix. Calculation of the emittance dilution

Moving the beam centroid as prescribed in Section 4 increases the total footprint occupied by the beam in the phase space over time. The increase can be characterized by an effective emittance $\varepsilon_1$:

$$\varepsilon_1^2 = \overline{(x_k + x_0)^2} \cdot \overline{(x_k' + x_0')^2} - \overline{(x_k + x_0)(x_k' + x_0')}^2 , \tag{A1}$$

where $x_k$, $x_k'$ are particle coordinates and angles with respect to the beam center; $x_0$ is the time-dependent position of the beam center from Eq.(20) and $x_0'$ is the angle of the beam center. It is calculated as a derivative of $x_0$ over the longitudinal coordinate

$$\begin{gathered} x_0 = C \sin(\omega t + \varphi_1) \\ x_0' = -\frac{C}{\beta_x}[\cos(\omega t + \varphi_1) + \alpha_x \sin(\omega t + \varphi_1)], \\ C \equiv \theta_1 \sqrt{\beta_x \beta_{x1}} \sin \varphi_x \end{gathered} \tag{A2}$$

where the $C$ is the oscillation amplitude. The bar in Eq.(A1) denotes averaging over both the ensemble and time. Averaging in Eq.(A1) gives

$$\varepsilon_1^2 = \left(\varepsilon_0 \beta_x + \frac{C^2}{2}\right)\left(\varepsilon_0 \frac{1+\alpha_x^2}{\beta_x} + \frac{C^2}{2\beta_x^2}(1 + \alpha_x^2)\right) - \left(\varepsilon_0 \alpha_x + \frac{C^2}{2\beta_x}\alpha_x\right)^2. \tag{A3}$$

After simplification of Eq.(A3), the effective emittance does not depend on the specific location where it is estimated:

$$\varepsilon_1 = \varepsilon_0 + \frac{C^2}{2\beta_x} = \varepsilon_0 + \frac{\theta_1^2 \beta_{x1} \sin \varphi_x^2}{2} \tag{A4}$$

The relative emittance increase depends quadratically on the oscillation amplitude

$$\frac{\varepsilon_1}{\varepsilon_0} - 1 = \frac{1}{2}\left(\frac{C}{\sigma_b}\right)^2. \tag{A5}$$